\documentclass[10.5pt]{IEEEtran}
\usepackage{graphicx}
\usepackage{amsmath}
\usepackage{multicol}
\usepackage{multirow}
\usepackage{stfloats}
\usepackage{booktabs}
\usepackage{mathrsfs}
\usepackage{enumerate}
\usepackage{amssymb}
\usepackage{algorithm}
\usepackage{algpseudocode}

\usepackage{array}
\usepackage{underscore}
\usepackage{breqn}
\usepackage{url}
\usepackage{bm}
\usepackage{subfigure}
\usepackage{float}
\usepackage{stfloats}

\usepackage{hyperref}
\hypersetup{hidelinks}
\usepackage{flushend}

\usepackage[square, comma, sort&compress, numbers]{natbib}
\usepackage{graphicx}
\usepackage{epstopdf}
\usepackage{caption}
\usepackage{diagbox}

\usepackage{caption}
\usepackage{color}

\usepackage[outerbars,color]{changebar}
\ifx\pdfoutput\undefined
\else\ifnum\pdfoutput>0
  \usepackage{pdfcolmk}
\fi\fi
\cbcolor{red}
\usepackage{fancyhdr}

\makeatletter
\newcommand{\Rmnum}[1]{\expandafter\@slowromancap\romannumeral #1@}
\makeatother

\usepackage{array}  \newcommand{\PreserveBackslash}[1]{\let\temp=\\#1\let\\=\temp}  \newcolumntype{C}[1]{>{\PreserveBackslash\centering}p{#1}}  \newcolumntype{R}[1]{>{\PreserveBackslash\raggedleft}p{#1}}  \newcolumntype{L}[1]{>{\PreserveBackslash\raggedright}p{#1}}

\ifCLASSINFOpdf
\else
\fi
\begin{document}

\title
{\huge
Wireless Power Transmitter Deployment for Balancing Fairness and Charging Service Quality
\huge}

%

\author{

Mingqing~Liu,
Gang Wang,~\IEEEmembership{\normalsize Member,~IEEE\normalsize},
Georgios B. Giannakis,~\IEEEmembership{\normalsize Fellow,~IEEE,\normalsize}
Mingliang~Xiong,\\
Qingwen Liu,~\IEEEmembership{\normalsize Senior Member,~IEEE\normalsize},
and Hao Deng

	\thanks{M. Liu, M. Xiong, Q. Liu and H. Deng are with the College of Electronic and Information Engineering, Tongji University, Shanghai 200000, China (e-mail: clare@tongji.edu.cn, xiongml@tongji.edu.cn, qliu@tongji.edu.cn, and dashena85754@qq.com).

G. Wang and G. B. Giannakis are with the Digital Technology Center and the Department of Electrical and Computer Engineering, University of Minnesota, Minneapolis, MN 55455, USA (e-mail: gangwang@umn.edu, georgios@umn.edu).}

}

\maketitle

\begin{abstract}
\normalsize
Wireless Energy Transfer (WET) has recently emerged as an appealing solution for power supplying mobile / Internet of Things (IoT) devices.
As an enabling WET technology, Resonant Beam Charging (RBC) is well-documented for its long-range, high-power, and safe ``WiFi-like'' mobile power supply.
To provide high-quality wireless charging services for multi-user in a given region, we formulate a deployment problem of multiple RBC transmitters for balancing the charging fairness and quality of charging service.
Based on the RBC transmitter's coverage model and receiver's charging / discharging model, a Genetic Algorithm (GA)-based and a Particle Swarm Optimization (PSO)-based scheme are put forth to resolve the above issue.
Moreover, we present a scheduling method to evaluate the performance of the proposed algorithms.
Numerical results corroborate that the optimized deployment schemes outperform uniform and random deployment in 10\%-20\% charging efficiency improvement.


\end{abstract}


\IEEEpeerreviewmaketitle

\section{Introduction}\label{Section1}

The IoT is featured with the ``seamless'' connectivity and communications of billions of smart devices, which provide different functionalities and serve personalized needs~\cite{wu2014cognitive}. Yet, the battery life of electronic devices is one of the current IoT dilemmas~\cite{2016Fadhel,IoTbattery,HuangC}. Moreover, with the rapid development of Mobile Internet (MI), the contradiction between battery life and power supply of mobile devices surges significantly. In this context, wireless charging, also known as WET, has recently emerged as an appealing solution for powering up mobile / IoT devices  \cite{mm2017wpt,ZhangR,Na2018Energy}.

In recent years, short-range WET based on inductive coupling and magnetic resonance has gained attention in certain range-limited (less than a meter) applications; see e.g., powering implanted medical devices \cite{2011medical}, as well as recharging smart phones \cite{2012Smartphone}. More recently, powering electronic devices in the far field (up to a few kilometers) using Radio Frequency (RF) signals has been shown to be feasible~ \cite{Shigeta2013Ambient,2014RF}. Nevertheless, these WET technologies are either challenged by the long distances, or facing difficulties in balancing high transfer power and safety \cite{lu2016wireless}.

\begin{figure}[t]
	\centering
    \includegraphics[scale=0.38]{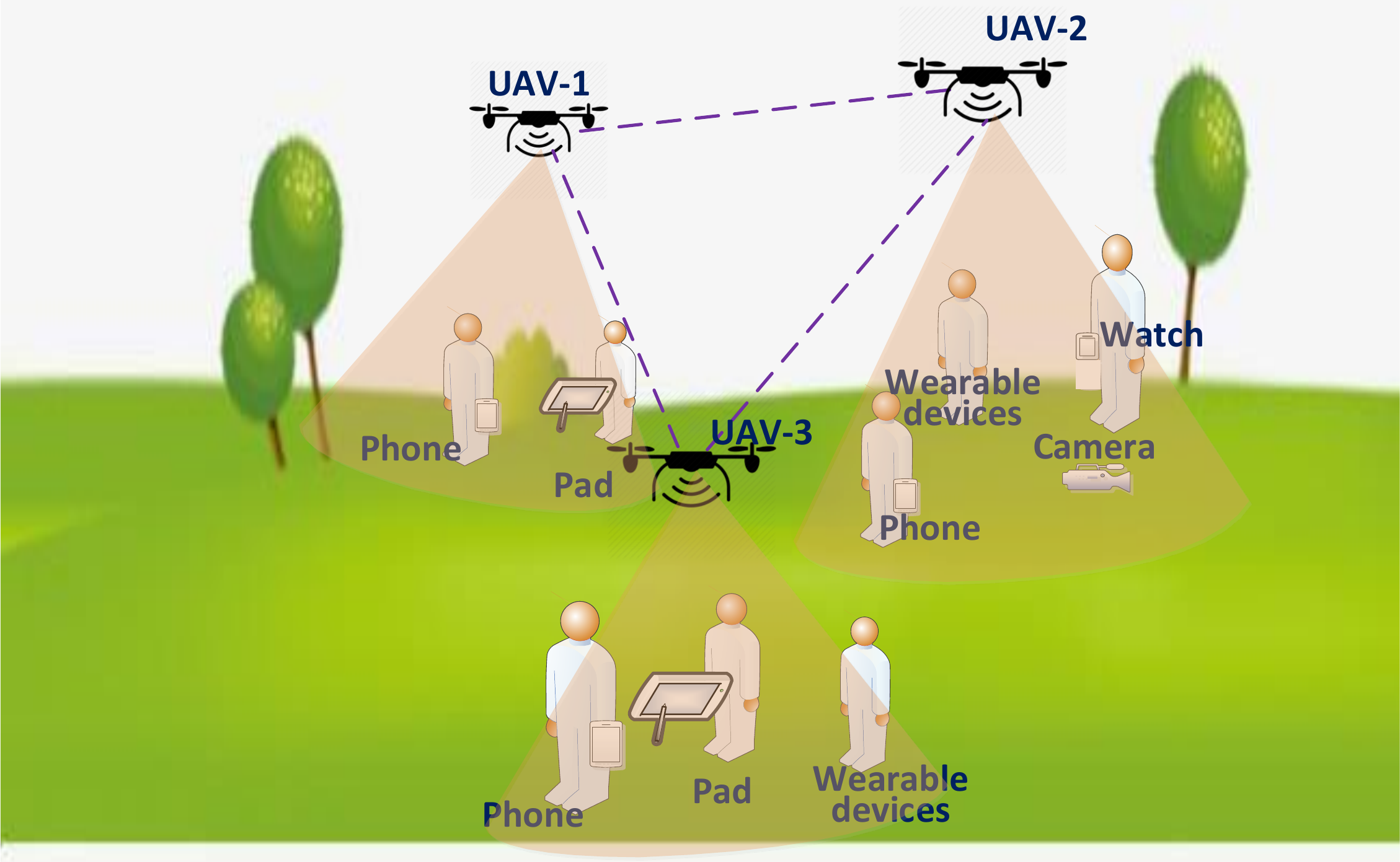}
	\caption{RBC application scenario.}
    \label{RBC Scenario}
    \end{figure}

 On the other hand, RBC, also known as Distributed Laser Charging (DLC), has been well-documented for its safe, high-power, and long-range wireless charging services \cite{liu2016dlc}. Besides, RBC can simultaneously charge multiple devices due to its broadcast property. Hence, RBC is well suited for providing far-field wireless power for mobile / IoT devices. See Fig. \ref{RBC Scenario} for the paradigm of an outside RBC application scenario, where the RBC transmitters are embedded in the Unmanned Aerial Vehicle (UAV) to provide wireless power for devices such as mobile phones, watches, and cameras, to name a few. Multiple UAV-based RBC transmitters work together to provide wireless charging services to a wide range of users. The RBC charging protocol is simple: devices embedded with an RBC receiver set up a WET connection with an RBC transmitter, receive energy transferred through invisible infrared light, and transform the beam power to electric power to charge batteries.

Similar to the signal attenuation in cellular transmission, there is power attenuation in WET \cite{BaseStation2016}. This is due to the fact that an RBC transmitter has a range of charging coverage, beyond which the transmitted power becomes negligible. Moreover, the charging profile, which means that the battery's request power varies during the charging process, needs to be followed to meet the user's charging service requirements. In diverse real-world situations,
we are often tasked with deploying RBC transmitters to provide wireless power for multi-user, with the consideration of charging fairness and quality of charging service. Thus, we formulate a deployment problem of RBC transmitters for optimally covering the electronic devices to be charged, and propose solutions based on RBC charging models and optimization algorithms.

Then, the RBC structure and the fundamental power attenuation principle are reviewed, relying on which an RBC coverage model is derived. Moreover, the charging / discharging model is introduced according to the battery's charging profile. Based on the above models, we develop a GA-based and a PSO-based approach searching for the optimal RBC transmitter deployment in a given area. Besides, we design a scheduling charging rule to evaluate the algorithm's performance.

This paper provides the deployment scheme for fair and high-quality wireless charging services. In order to realize wireless power supply anywhere and anytime, our contributions are as follows:

1) With the consideration of the RBC transmitter's coverage model and charging / discharging model, we first formulate the deployment problem of RBC transmitters to provide wireless charging services for users in a wide range, balancing the charging fairness and quality of charging service.

2) To cope with the above issues, we propose two optimal deployment schemes for RBC transmitters based on GA and PSO, respectively. Numerical results showcase our proposed deployment schemes outperform the uniform and random deployment method.

The remainder of this paper is as follows. Section \ref{Section2} outlines basic models of an RBC system, in which a coverage model for a single RBC transmitter, an RBC receiver charging / discharging model, and a scheduling rule are presented. We formulate the problem discussed in this paper in Section \ref{Section3}.
A GA-based and a PSO-based scheme for RBC transmitter deployment are developed in Section \ref{Section4}.
Performance of the proposed deployment schemes are evaluated in Section \ref{Section5}. Finally, the paper is concluded in Section \ref{Section6}.


\section{RBC Model}\label{Section2}

\subsection{RBC structure and features}\label{}

\begin{figure}[t]
	\centering
    \includegraphics[scale=0.5]{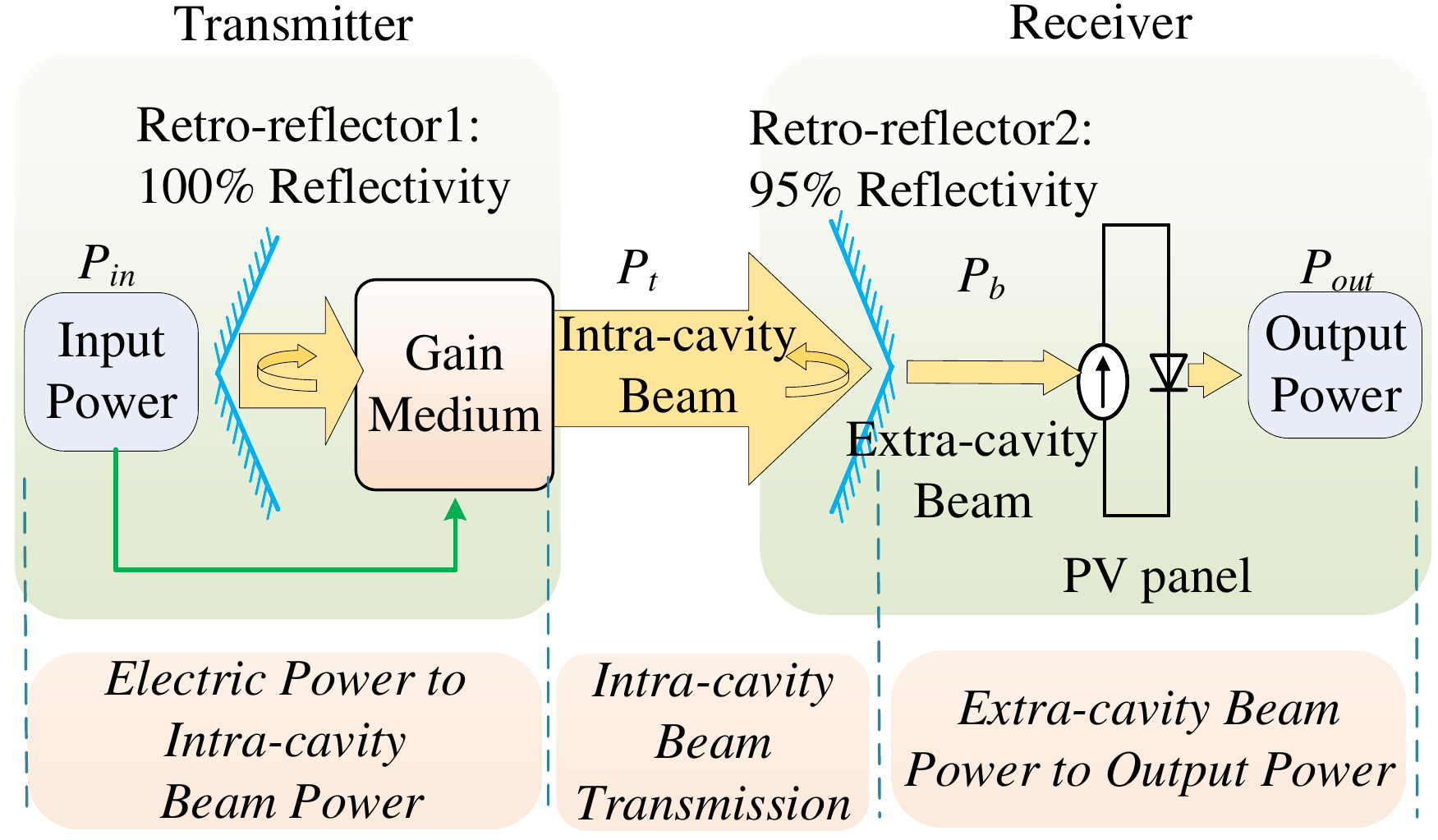}
    \vspace{.1em}
	\caption{RBC system structure.}
    \label{RBC Structure}
    \end{figure}

RBC is a novel WET technology, which aims at providing high-power, long-range, and safe wireless charging services \cite{liu2016dlc}. In contrast to the traditional laser charging schemes that transfer energy through laser emitted from integrated devices, RBC transfers energy through an intra-cavity resonant beam with separated cavity structure \cite{zhang2018distributed2}. See Fig. \ref{RBC Structure} for a demonstration of a typical RBC system, which comprises two main components: a transmitter and a receiver. The RBC transmitter and receiver jointly form a stable resonant cavity, in which the power is transferred from the transmitter to the receiver through a resonant beam shuttling back and forth. The input power at the transmitter side first stimulates the gain medium to generate an intra-cavity resonant beam. The intra-cavity resonant beam oscillates in the stable cavity between retro-reflector 1 (of $100\%$ reflectivity) and retro-reflector 2 (of $95\%$ reflectivity). Subsequently, the intra-cavity beam partially passes through the partially reflective retro-reflector 2 at the receiver side to form an extra-cavity beam. Finally, the extra-cavity beam at the receiver side is converted into the output electrical power using a photovoltaic (PV) panel, which can be readily used to charge electronic devices.

The distributed structure of RBC can guarantee safe charging of multiple devices simultaneously, while meeting high-power transmission requirements. Any foreign object entering the line of sight between the transmitter and the receiver (i.e., the resonant cavity) blocks the path of photons, thus breaking the conditions of a stable cavity, and automatically cutting off the resonant beam. Even in unexpected situations, the high-power beam will have already been terminated before it causes any damage. In this regard, RBC is intrinsically safe. Moreover, according to the characteristics of the retro-reflectors, the RBC transmitter and receiver can form a resonant cavity over a wide-angle range. Then alignment lines are automatically generated requiring no specialized alignment devices. This property enables RBC with mobility.

Besides the aforementioned merits, RBC is also featured with concurrent charging, hot-spot charging, compact size, electromagnetic interference-free, as well as wavelength-agnostic; see \cite{liu2016dlc, zhang2018distributed2, fang2018, Qing2017,Qing2019} for a recent survey and related discussions on RBC.

\subsection{RBC transmitter coverage model}\label{Section2B}
From Fig.~\ref{RBC Structure} and \cite{zhang2018distributed2}, the WET in RBC system is divided into three stages: i) the input electric power $P_{in}$ to the intra-cavity beam power $P_{t}$; ii) the intra-cavity beam transmission, after which the $P_{t}$ turns into extra-cavity beam power $P_{b}$; and, iii) the extra-cavity beam power $P_{b}$ to the output power $P_{out}$. Suppose the conversion efficiency of the input power $P_{in}$ to intra-cavity beam power $P_t$ is $\eta_{t}$, the conversion module of stage i) is depicted as~\cite{zhang2018distributed2}
\begin{equation}\label{pinpt}
    P_{t} = \eta_{t}P_{in}.
\end{equation}

Similar to the fact that transmitted signals get attenuated in a communication system, the power transferred in an RBC system gets attenuated with the increase of transmission distance as well. Thus, the extra-cavity beam power $P_b$ after stage ii) can be found as
\begin{equation}\label{beam-power}
    P_b = f(d) P_t + C
\end{equation}
where $C>0$ is a constant depending on parameters of the system, and $f(d)$ is an attenuation function in regards to the transmission distance $d$ between the transmitter and the receiver, which can be depicted as
\begin{equation}\label{f-d}
    f(d) = \frac{2(1-R)m}{(1+R)(\delta - {\rm lnR})}
\end{equation}
where $R$ is the reflectivity of the output mirror at the receiver, $m$ denotes the ratio of gain medium diameter to aperture, and $\delta$ is the diffraction loss in the resonant cavity as following
\begin{equation}\label{diffraction-loss}
    \delta = e^{-2 \pi {\frac{a^2}{\lambda (l+d)}}},
\end{equation}
where $a>0$ is the shared radius of the two retro-reflectors in RBC, $\lambda$ the wavelength of the beam, $l>0$ is the distance between the gain medium and retro-reflector 1, and $d$ denotes the transmission distance between the transmitter and the receiver.

Based on the previous work of \cite{zhang2018distributed2} and \cite{WWang2018}, we can write the RBC output power as a function of the input power and the transmission distance, namely
\begin{equation}\label{pout}
    P_{out}=\alpha (f(d) \eta_{t}P_{in} + C)+\beta,
\end{equation}
where $\alpha$ and $\beta$ are constant coefficients obtained by the fitting process in~\cite{zhang2018distributed2}.

Given a certain input power $P_{in}$, the output power $P_{out}$ declines as the transmission distance $d$ increases due to the diffraction loss, which can be described as RBC power attenuation with the augment of $d$. Based on the RBC energy attenuation model in \eqref{f-d}, \eqref{diffraction-loss}, and \eqref{pout}, we can obtain the formula for transmission distance $d$ as
\begin{equation}\label{distance2}
    d = M \ln^{-1}\left(\frac{K P_{in}}{Z P_{out} + U} + N\right)-l
\end{equation}
where the constants $M:=-2\pi a^2/\lambda$, $K:=2m(1-R)\alpha \eta_t$, $Z:=1+R$, $U:=-(1+R)(\beta+\alpha C)$, $N:=\ln R$, and $l$ is the distance between the gain medium and the retro-reflector 1.

\begin{table}[!t]
    \setlength{\abovecaptionskip}{0pt}
    \setlength{\belowcaptionskip}{5pt}
    \centering
        \caption{~Parameters of the Graphical Coverage Model}
        \vspace{.7em}
    \begin{tabular}{C{1.5cm} C{1.8cm}|C{1.5cm} C{2cm}}
    \hline
     \textbf{Parameters} & \textbf{Value} & \textbf{Parameters} & \textbf{Value} \\
    \hline
    \bfseries{$a$} & { $1.5$ } & { $m$ } & { $0.8$ } \\
    \bfseries{$\lambda$} & { $1.064\times 10^{-3}$ } & { $\eta_t$ } & { $0.2849$ } \\
    \bfseries{$C$} & { $-5.64$ } & { $\alpha$ } & { $0.3487$ } \\
    \bfseries{$\beta$} & { $-1.535$ } & { $\pi$ } & { $3.14$ } \\
    \bfseries{$R$} & { $0.88$ } & { $l$ } & { $0$ } \\
    \bfseries{$h$} & { $3$m/$5$m  } & { $P_{in}$ } & { $150$W/$200$W } \\
    \hline
    \label{parameters}
    \end{tabular}
    \end{table}

To practically deploy an RBC network, it is instrumental to understand and predict the coverage of an RBC transmitter, that is, to determine the radius (i.e., covering range $r$)  of a transmitter's coverage. In this paper, all receivers are assumed to be placed at the same plane. The coverage of an RBC transmitter is defined as an area on the plane where the received power by receivers is above a certain level. In the ideal case, given a required minimum received power $P_{min}$, the coverage of an RBC transmitter on the plane is a circle, where the receivers' output power $P_{out}\ge P_{min}$.

Suppose the RBC transmitter is placed at a height of $h$, the relationship between $r$ and $d$ following the Pythagorean theorem can be depicted as
\begin{equation}\label{radius}
    r = \sqrt{d^2 - h^2}.
\end{equation}

Thus, the covering range can be obtained as
 \begin{equation}\label{coveringmodel}
 \begin{split}
    r = \sqrt{\left[M \ln^{-1}\left(\frac{K P_{in}}{Z P_{out} + U} + N\right)-l\right]^2 - h^2},
    ~~ P_{out}\ge P_{min}
 \end{split}
\end{equation}

The RBC coverage model derived from \eqref{coveringmodel} suggests that, when the input power, the transmitter height, and the minimum acceptable charging power $P_{min}$ are known, the covering range $r$ of an RBC transmitter can be readily computed.

\begin{figure}[!t]
	\centering
    \includegraphics[scale=0.57]{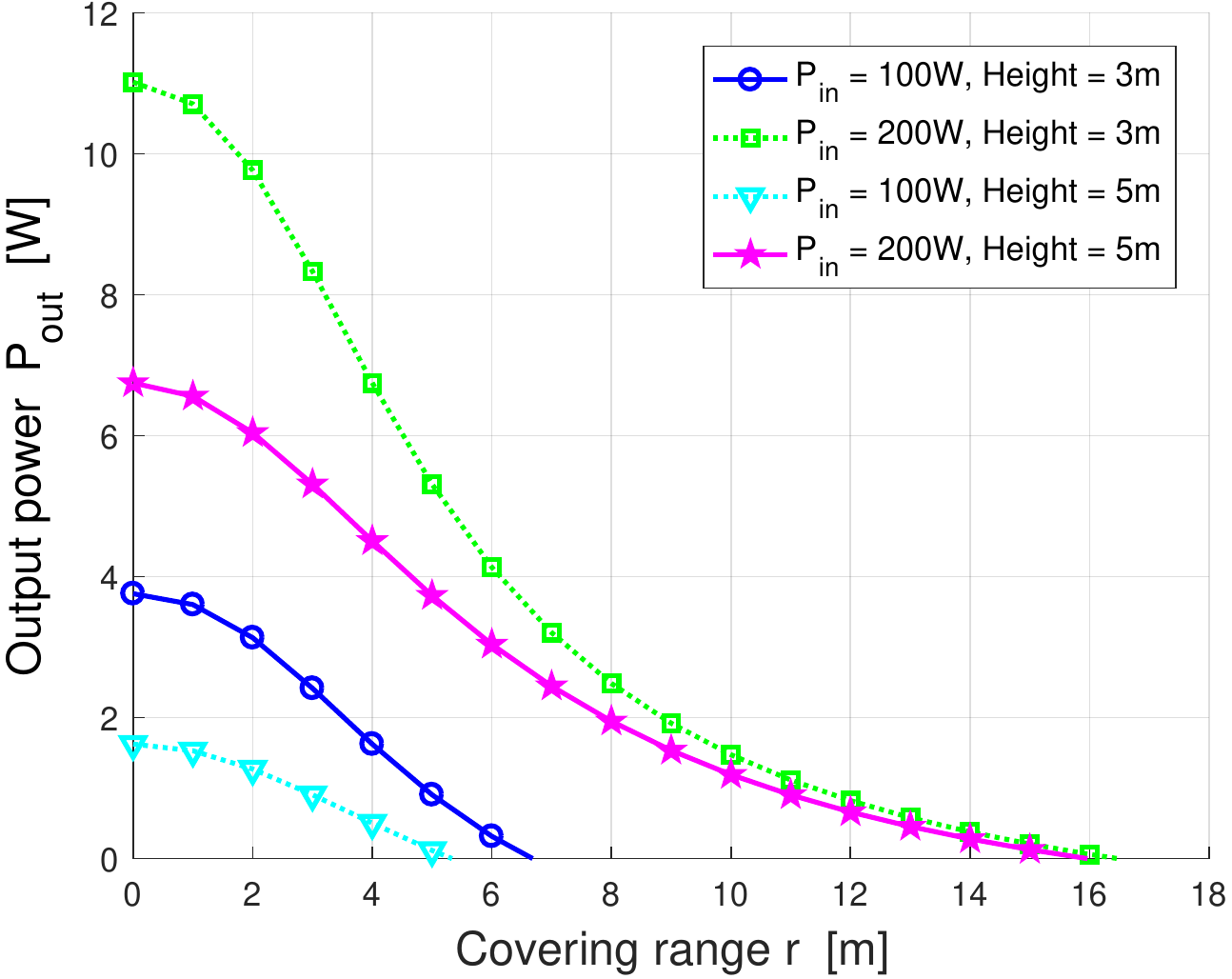}
	\caption{Output power as a function of the covering range.}
    \label{2dimension}
    \end{figure}

To facilitate, we plot the output power $P_{out}$ as a function of the covering range $r$ in Fig. \ref{2dimension}, along with the parameter values in \eqref{coveringmodel} given in Table \ref{parameters}. Figure \ref{2dimension} depicts how the output power of an RBC system varies as the covering range grows when the input power is $100$W and $200$W, and the transmitter height is $3$m and $5$m, respectively. Evidently, when the RBC transmitter is placed at a height of $3$m and the input power is $200$W, it can supply more than charging power of $5$W to all receivers in the coverage within a covering radius of $5$m. If a higher transmitter height and a larger service covering range are necessary, more input power is required.

%

\subsection{RBC receiver charging and discharging model}\label{}
To better satisfy the charging request of mobile device users, we introduce the charging / discharging model to simulate the actual charging process. According to the battery's charging profile, mobile device batteries are charged with variable current and voltage rather than fixed current and voltage during the charging procedure. That is, under different SOC, the power that the battery requests differs. Thus, in order to provide receivers with their preferred charging power at any state, we should at first determine the charging mode. The RBC charging model depicts the relationship between the preferred charging power and the remaining capacity of the battery.

\begin{figure}[!t]
	\centering
    \includegraphics[scale=0.575]{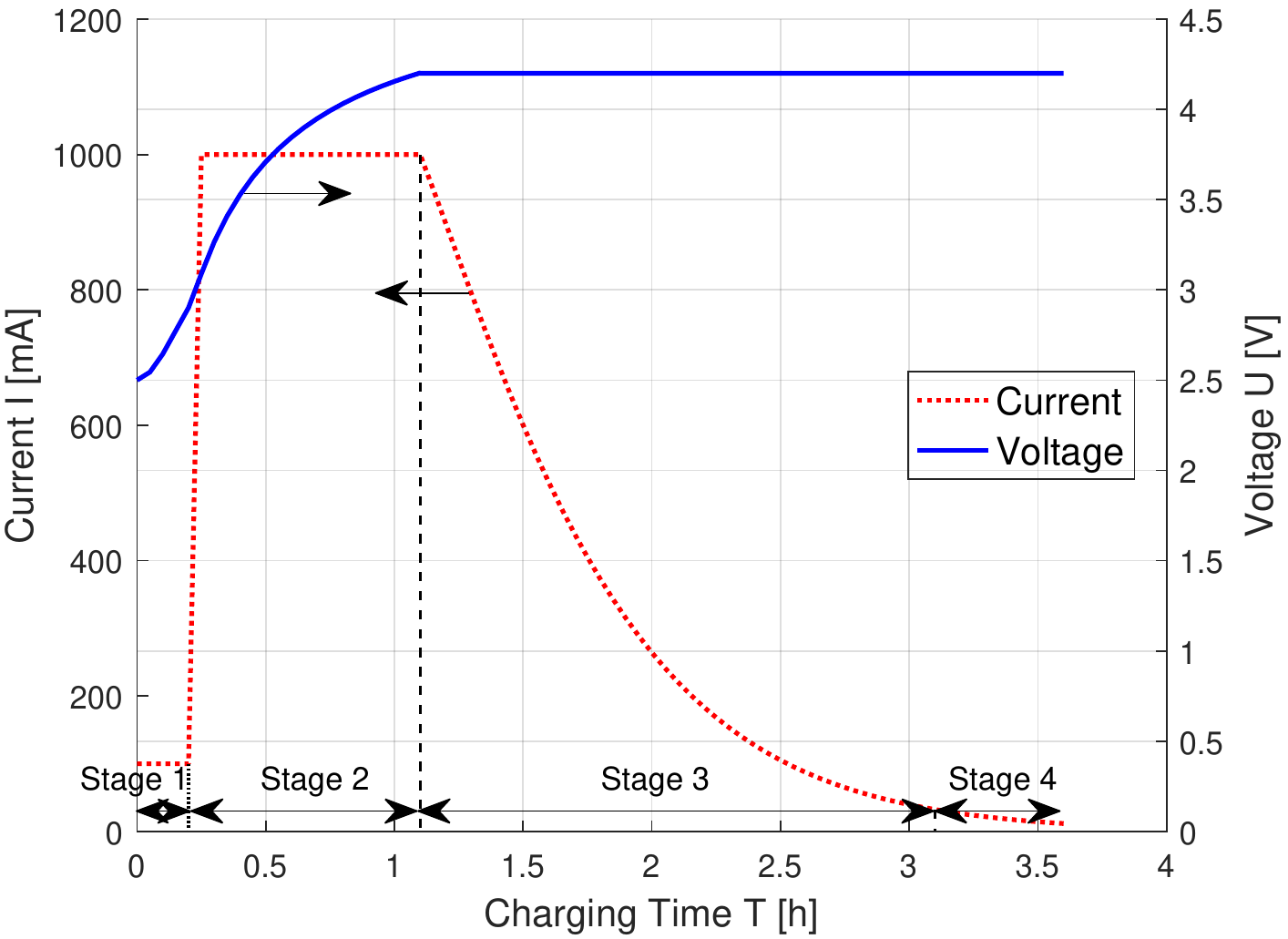}
    \vspace{.3em}
	\caption{Charging profile of lithium-ion battery.}
    \label{chargingprofile}
    \end{figure}

We adopt a lithium-ion battery as an analysis model, which is the most widely used battery in mobile devices. Figure.~\ref{chargingprofile} depicts the charging profile of the $4.2$V/$1$A, $1000$mAh lithium-ion battery. The charging process can be divided into four stages~\cite{chargingprofile1}. In stage 1 which is called as trickle charging, the charging voltage is below $3$V while the charging current is around $100$mA; in stage 2, the charging current is constant at $1000$mA, and the charging voltage increases to $4.2$V; in stage 3, the battery preferred to be charged with a constant voltage as $4.2$V, and with a decreasing current from $1000$mA to $40$mA; and in stage 4, the battery is almost fully charged,as a result of which the charging current is below $40$mA.

Based on the charging profile, a fitting function is configured to fit the relationship between the residual energy and the preferred charging power~\cite{fwfair}:

\begin{equation}\label{fittingfunction}
    P_c(x)=\frac{\beta_{1}^{\prime} x^{4}+\beta_{2}^{\prime} x^{3}+\beta_{3}^{\prime} x^{2}+\beta_{4}^{\prime} x+\beta_{5}^{\prime}}{x^{5}+\alpha_{1}^{\prime} x^{4}+\alpha_{2}^{\prime} x^{3}+\alpha_{3}^{\prime} x^{2}+\alpha_{4}^{\prime} x+\alpha_{5}^{\prime}},
\end{equation}
where $x$ represents the current battery remaining energy capacity, $P_c$ represents the preferred charging power of the battery, and the coefficients are listed in Table~\ref{coefficients}.

\begin{table}[!t]
    \setlength{\abovecaptionskip}{0pt}
    \setlength{\belowcaptionskip}{5pt}
    \centering
        \caption{~Capacity-preferred power fitting function coefficients}
        \vspace{.7em}
    \begin{tabular}{C{1.5cm} C{1.8cm}|C{1.5cm} C{2cm}}
    \hline
     \textbf{Coefficients} & \textbf{Value} & \textbf{Coefficients} & \textbf{Value} \\
    \hline
    \bfseries{$\beta_{1}^{\prime}$} & { $-21.65$ } & { $\alpha_{1}^{\prime}$ } & { $-10.7$ } \\
    \bfseries{$\beta_{2}^{\prime}$} & { $141.2$ } & { $\alpha_{2}^{\prime}$ } & { $41.01$ } \\
    \bfseries{$\beta_{3}^{\prime}$} & { $-11.5$ } & { $\alpha_{3}^{\prime}$ } & { $-1.509$ } \\
    \bfseries{$\beta_{4}^{\prime}$} & { $0.1526$ } & { $\alpha_{4}^{\prime}$ } & { $-0.3997$ } \\
    \bfseries{$\beta_{5}^{\prime}$} & { $0.008358$ } & { $\alpha_{5}^{\prime}$ } & { $0.0362$ } \\
    \hline
    \label{coefficients}
    \end{tabular}
    \end{table}

    Besides, in practice, the mobile devices are probably being used during the charging process, which means that the receivers will also discharge while being charged. Thus, we add the discharging model according to the probabilities and discharging power values of mobile phone applications which may be used~\cite{fwfair}.

    The discharging power of a receiver $P_d$ can be depicted as~\cite{fwfair}:
    \begin{equation}\label{discharge}
P_{d}=\operatorname{randsrc}\left(1,1,\left[\bm {P}_{u} ; \bm{U}_{p}\right]\right),
\end{equation}
$\operatorname{randsrc}(m,n,[alphabet; prob])$  generates an m-by-n matrix, each of whose entries is independently chosen from the entries in the row vector alphabet. The row vector prob lists corresponding probabilities, so that the symbol $alphabet(j)$ occurs with probability $prob(j)$, where $j$ is any integer between one and the number of columns of alphabet.  $\bm{U}_p$ represents usage rates of five representative working statuses of mobile phones, i.e., standby, video, social software (e.g., Facebook, Twitter), game
and music, which is:
\begin{equation}
\bm{U}_{p}=\{28.39 \% \text { } 12.35 \% \text { } 24.69\%\text { }  12.35 \% \text { } 22.  22 \%\},
\end{equation}
and $\bm{P}_u$ represents the corresponding discharging power of each working status, as shown as:
\begin{equation}
\bm{P}_{u}=\{0.0076~0.4289~0.4348~0.6766~0.1706\}.
\end{equation}

\section{Multi-transmitter deployment modeling}\label{Section3}
\subsection{Problem formulation}\label{}
It is evident from the RBC transmitter coverage model that an ordinary RBC transmitter provides only limited power, which may be sufficient for charging low-power electronic devices such as mobile phones placed within a range of $2\sim5$m. To supply power for high-power devices placed on any point of the plane in a certain area, one may require multiple transmitters. Referring to the deployment of base stations in mobile communications, two factors should be considered in the deployment of RBC transmitters: charging fairness and quality of charging service. To balance the above factors, we streamline the problem to the following two subgoals:

1) as many receivers as possible can be covered by the RBC transmitters' wireless charging services. According to the RBC transmitter coverage model, one transmitter covers limited receivers. Thus, to guarantee the fairness of charging in an area, a certain number of transmitters should be optimally deployed to cover more receivers;

2) covering as many receivers with less remaining capacity as possible. The charging demand of the receiver remaining less energy is more urgent, and it will request more power according to the charging model. Therefore, to meet most urgent needs of users, the sum of the remaining power of the covered receivers should be as small as possible.

Correspondingly, given a set of to-be-covered receivers' coordinates $\bm{R} = \left\{\mathbf{r}_{1}, \mathbf{r}_{2}, \ldots, \mathbf{r}_{N_{total}} \right\}$, and a candidate set of deployed transmitters' coordinates $\bm{T}_c = \left\{\mathbf{t}_{1}, \mathbf{t}_{2}, \ldots, \mathbf{t}_{N_t} \right\}$, we can construct an objective function as follows:
\begin{equation}\label{objectivesum}
Q (\bm{T}_c) = \omega_1Q_1+\omega_2Q_2
\end{equation}

\begin{equation}
Q_1(\bm{T}_c) = \max\left\{\frac{N_{covered}}{N_{total}}\right\}
\end{equation}

\begin{equation}\label{objective1}
Q_2(\bm{T}_c) = \max\left\{ \frac{\sum_{i=1}^{N_{covered}}(1-S_c(\mathbf{r}_i))}{\sum_{i=1}^{N_{total}}(1-S_c(\mathbf{r}_i))}\right\}
\end{equation}
where $N_{total}$ is the number of total receivers, $N_t$ is the number of deployed RBC transmitters, and $S_c(\mathbf{r}_i)$ is the remaining capacity percentage of each receiver, which can be defined by the ratio of residual energy $E_r(\mathbf{r}_i)$ and the total battery energy $E_{total}$ of each receiver:
\begin{equation}\label{SOC}
    S_c(\mathbf{r}_i) = \frac{E_r(\mathbf{r}_i)}{E_{total}(\mathbf{r}_i)}\times 100\%.
\end{equation}

 $N_{covered}$ is the number of receivers covered by the RBC transmitters, which can be depicted as

%
\begin{equation}
  N_{covered}=\sum_{i=1}^{N_{total}} \operatorname {bool }\left(\sum_{k=1}^{N_t} \operatorname {bool } \left(\left\|\mathbf{t}_{k}-\mathbf{r}_{i}\right\|_{2} \le r\right)>0\right),
\end{equation}
where
\begin{equation}
\operatorname{bool}(\text {Statement})=\left\{\begin{array}{ll}{1} & {\text {, Statement is true; }} \\ {0} & {\text {, Statement is false. }}\end{array}\right.
\end{equation}
 $\mathbf{t}_k $ is the coordinate of k-th transmitter in $\bm{T}_c$, $\mathbf{r}_i $ is the coordinate of i-th receiver in $\bm{R}$, and $r$ is the covering range of an RBC transmitter. $\omega_1, \omega_2\in [0,1]$ in \eqref{objectivesum} are weights of objective functions $Q_1$ and $Q_2$, which represent the balance factor of the charging fairness and the quality of charging service, respectively.

With the aforementioned models, our problem can be stated as: Given a 2-D area $\bm{G}$ with multiple to-be-charged mobile devices, based on the RBC coverage model and charging / discharging model, how to find a RBC transmitter placement that can satisfy the objective function optimally. Then we formulate our problem as follows: find a set of transmitter coordinates $\bm{T}^{\star} = \left\{ \mathbf{t}_{1}, \mathbf{t}_{2}, \ldots, \mathbf{t}_{N_t}\right\}$ which can maximize the objective function $Q$, that is

\begin{equation}
\bm{T}^{\star} =  \mathop{\arg\max}_{ \bm{T}_c \in \bm{J}}~Q(\bm{T}_c),
\end{equation}
where $\bm{J}$ is the union of all possible sets of deployed transmitters generated in the region, which can be defined as ($\bm{N}_+ $ represents
the set of all positive integers)
\begin{equation}
\begin{aligned}
\bm{T} &= \left\{(x,y)_k ~|~ (x,y)_k \in \bm{G}, 1\le k\in \bm{N}_+\le N_t \right\}\\
\bm{J} &= \left\{ \forall~\bm{T}_c ~|~ \bm{T}_c\in \bm{T} \right\}.
\end{aligned}
\end{equation}

\subsection{Evaluation method}\label{Evaluation}
To evaluate the performance of the optimized placement method, we establish two evaluation indexes as follows:

 1) maximizing the object function with least transmitters, which is defined as the quality of coverage;

 2) maximizing the average residual capacity of all receivers in the region after charging for a certain time, which is defined as the charging efficiency.

 Hence, we set a scheduling rule for the deployed RBC transmitters to charge the receivers within their coverage. The scheduling process is described below:

1) Attribution classification of receivers: one receiver may be covered by multiple RBC transmitters, while at the same time one receiver should only be charged by one transmitter for simplicity. Thus, we at first classify which transmitter that the receiver is belonged to according to the distance between them. That is, one receiver is only assigned to the transmitter closest to it, and will only be charged by this transmitter.

2) Charging and discharging: based on the charging and discharging model, we at first divide the total charging time into many time slots. In each time slot, all receivers belonging to a certain transmitter will charge and discharge referring to equations \eqref{fittingfunction} and \eqref{discharge}, except for that the receivers are fully charged.

In details, the receivers have different residual energy $E_r(\mathbf{r}_i)$ at the initial state, according to which the receivers will request various power $P_c(\mathbf{r}_i)$. Moreover, the mobile device will also discharge $P_d(\mathbf{r}_i)$ during the charging process. Under a certain deployment, the residual energy of each receiver covered by the transmitter at the time of $t$ will change after a time slot $t_d$ as
\begin{equation}\label{objective1}
E_r(\mathbf{r}_i,t + t_d) = E_r(\mathbf{r}_i, t) + P_c(\mathbf{r}_i)\times t_d - P_d(\mathbf{r}_i)\times t_d.
\end{equation}

After charging for a certain time, the average residual capacity of the total receivers in the whole region will be increased with the optimized deployment scheme.

\subsection{Receiver distribution model}
It's not in line with reality to assume that all users carrying devices to be charged are uniformly distributed in a certain area. Therefore, we draw on a non-uniform distribution generally used by base station deployment in mobile communication networks - Thomas distribution to simulate the distribution of wireless charging service users~\cite{userdistribution1, userdistribution2}.

In the Thomas distribution, there are two kinds of nodes which can both represent the receivers: cluster head nodes and common nodes. Cluster head nodes are central nodes, of which the distribution obeys the Poisson distribution, and the uniform density is $\lambda$. Common nodes surrounding the central nodes within a circle of certain radius follows the Gaussian distribution, of which the mean is $\mu$. $\mu$ is a
variable determined by the central nodes which are generated through the Poisson distribution. The covariance representing the radius of circle centered on the cluster head node is $\sigma$~\cite{userdistribution3}.
Thus, for intuitive presentation, we define the Probability Density Function (PDF) of Thomas distribution with two parameters as $T_m(\lambda, \sigma)$.

\section{Multi-transmitter Deployment algorithm}\label{Section4}


Determining the RBC transmitter deployment is tantamount to finding a set of placement position coordinates in a certain region.
GA and PSO are two commonly used algorithms to generate high-quality solutions for optimization and search problems \cite{GAcoverage2008,PSOWSN1}, which has been adopted to deal with the nodes deployment in the Wireless Sensor Networks (WSN) \cite{2013Yoon,PSOWSN2}. Hence, we solve the optimal transmitter deployment problem based on GA and PSO, respectively.

In this section, we will develop a GA-based and a PSO-based deployment for providing optimal wireless charging services in a given region. The design and implementation of the two proposed deployment schemes are presented.

In this paper, we suppose the region to be covered is a rectangular area with a length of $L$ and a width of $W$. The optimization objective of both the GA-based and PSO-based charging efficiency-oriented deployment algorithm is to obtain a set of transmitters' location coordinates for maximizing the objective function \eqref{objectivesum} with least transmitters. Assuming the region to be covered is a rectangular of area $\bm{G}$, then the coverage region is a set of cell's coordinates which can be presented as
 \begin{equation}\label{discrete}
   \bm{G} =  \left\{(x, y)~|~ 0\le x\le L, 0\le y\le W\right\}.
\end{equation}

As mentioned in the receiver distribution model, we assume that the receivers are scattered non-uniformly in a given region, obeying the Thomas distribution. Thus, the coordinates of receivers in $\bm{R} = \left\{\mathbf{r}_{1}, \mathbf{r}_{2}, \ldots, \mathbf{r}_{N_{total}} \right\}$ follows the condition below
\begin{equation}\label{}
    \forall \mathbf{r}_i \in \bm{R},~\mathbf{r}_i \thicksim T_m(\lambda,\sigma).
\end{equation}

\subsection{GA-based deployment}\label{}

GA starts with a set of randomly generated candidate solutions (a.k.a. the population) to obtain an optimal solution through bio-inspired operators such as mutation, crossover, and selection \cite{GA2005}. Each candidate solution (a.k.a. individual) in the population has a set of properties (a.k.a. the chromosome) which can be mutated and altered. The evolution is an iterative process with the population in each iteration called a generation. In each generation, the fitness of every individual usually defined to be the value of the objective function is evaluated. Then some individuals with larger fitness are selected to breed a new generation population through a combination of crossover and mutation. The new generation consisting of candidate solutions are to be used in the next iteration. After several iterations, the algorithm converges to the best individual.

In GA-based transmitter deployment, each individual $\bm{P}_i$ represents a possible deployment solution containing series random transmitter's coordinates, which can be described as
\begin{equation}\label{individual}
    \bm{P}_i =\{(x_{ij}, y_{ij})~|~1 \le i \le N_p, 1 \le j \le N_t, (x_{ij}, y_{ij})\in \bm{G}\}
\end{equation}
where $i,j \in \bm{N}_+$,  $(x_{ij}, y_{ij})$ is the coordinate of each transmitter, and $x_{ij}$, $y_{ij}$ are generated randomly within the border of the coverage region. A certain number of individuals $N_p$ consist of the population, and $N_t$ is the number of transmitters contained in one individual.
The optimal solution is a set of transmitters' coordinates maximizing the objective function \eqref{objectivesum} (i.e., the best individual).

Each receiver is initialized with a random battery remaining capacity percentage, according to \eqref{SOC}. In this case, we adopt a $4.2$V/$1$A, $1000$mAh lithium-ion battery, of which the total battery energy is $6.3865$J. Thus, we set the random remaining capacity of each receiver between $0.1-0.9$. Then, according to \eqref{objectivesum}, for any candidate set of transmitter coordinates $\bm{P}_i$, the fitness function of the GA-based deployment can be formed as follow
\begin{equation}\label{fitness}
\begin{aligned}
    g(\bm{P}_i) = \omega_1\times\frac{N_{covered}}{N_{total}} + \omega_2\times\frac{\sum_{i=1}^{N_{covered}}(1-S_c(\mathbf{r}_i))}{\sum_{i=1}^{N_{total}}(1-S_c(\mathbf{r}_i))}.
\end{aligned}
\end{equation}

The optimal RBC transmitter deployment can be found by searching for a set of coordinates to maximize the objective function \eqref{objectivesum} with GA. Algorithm 1 shows the execution flow of the optimal transmitter deployment search through GA.


The initialization consists of three parts:
\begin{enumerate}
 \item[\bf P1)]Set the number of RBC transmitters $N_t$, and set the maximum number $N_p$ of individuals in the population. Generate a population containing $N_p$ individuals and the coordinate of each individual $\bm{P}_i$ in the population is generated randomly within the border of the region. That is, $x_{ij}$ and $y_{ij}$ in \eqref{individual} are random floating values in $[0, L]$ and $[0, W]$.

\item[\bf P2)]Compute the fitness value for each individual using \eqref{fitness}. According to \eqref{coveringmodel}, the covering range $r$ of each RBC transmitter can be determined. If the distance between any transmitter coordinate in an individual and a receiver's coordinate in the region is less than $r$, this receiver is meant to be covered. All the receivers that can be covered by an individual in the region form the $N_{covered}$ of this individual. Then, using \eqref{fitness}, the fitness value for each individual is obtained.

\item[\bf P3)]Select the individual with the largest fitness value among the population and save it in the last position of the population array.
\end{enumerate}

\begin{algorithm}[!h]
    \caption{GA-based deployment}
    \begin{algorithmic}[1]
    \Require $L$, W, $r$;
    \State \text{Initialization:}
    \State $\text{\  Initializing\ the\ population}$;
    \State $\text{\  Compute\ the\ fitness\ value\ with\ \eqref{fitness}}$;
    \State $\text{\  Select\ the\ individual\ with\ the\ largest\ fitness}$;
    \While {$Iterator \le It_{max}$}
    \State $\text{\ Select\ the\ parents\ with\ crossover\ rights}$\ (cf. S1);
    \State $\text{\ Crossover\ and\ produce\ the\ new\ offspring}$\ (cf. S2);
    \State $\text{\ Individual\ mutation}$\ (cf. S3);
    \State $\text{\ Compute\ each\ individual's\ fitness\ with\ \eqref{fitness}}$\ (cf. S4);
    \State $\text{\ Optimize\ the\ best\ and\ worst\ individuals}$\ (cf. S5);
    \EndWhile
    \State \Return{$\text{The\ best\ individual}$}.
    \label{TDGA}
    \end{algorithmic}
    \end{algorithm}

\begin{figure}[!t]
	\centering
    \includegraphics[scale=0.42]{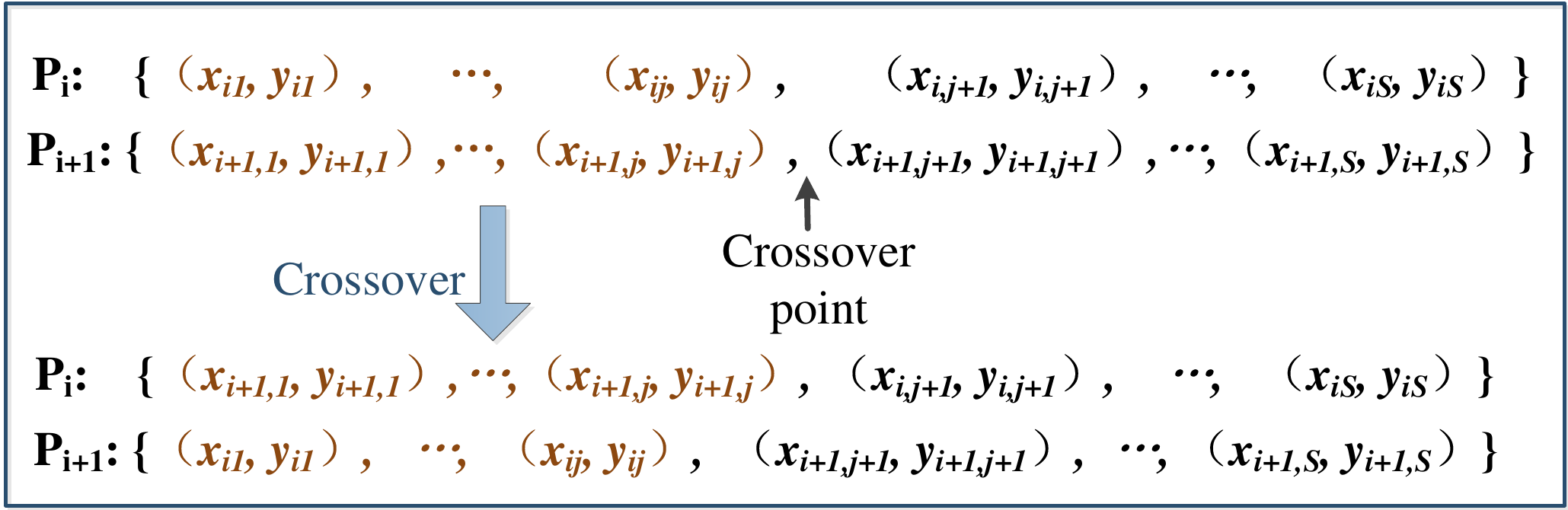}
	\caption{Crossover process of two individuals.}
    \label{crossover}
    \end{figure}

Steps of our GA-based deployment in Algorithm 1 are summarized as follows:

\begin{enumerate}
\item[\bf S1)]\textbf{Selection:} In each generation, a portion of the individuals in the current population are selected to breed a new generation. Candidate solutions (i.e., individuals) with larger fitness value, as measured by \eqref{fitness} are more likely to be selected. We adopt the roulette wheel selection to select individual solutions \cite{RWS2016}, in which the probability of each individual being selected is proportional to its fitness value. If the population size is $N_p$, and the fitness value of individual $\bm{P}_i$ is $g(\bm{P}_i)$, then the selection probability $P(\bm{P}_i)$ of $\bm{P}_i$ is

\begin{equation}\label{probality}
    P(\bm{P}_i) = \frac{g(\bm{P}_i)}{\sum_{i=1}^{N_p} g(\bm{P}_i)}.
\end{equation}

\item[\bf S2)]\textbf{Crossover:} Set a number $p_c$ in [0,1] as the crossover probability at first. Randomly select the crossover point, and the crossover of every two adjacent individuals carries out at the point if meeting the crossover probability. Figure \ref{crossover} shows the crossover process of two individuals. The coordinates before the crossover point of the two individuals exchange locations with each other while the coordinates after the crossover point stay the same.

\item[\bf S3)]\textbf{Mutation:} Set a number $p_m$ in [0,1] as the mutation probability at first. Randomly change coordinate value of each RBC transmitter in the individual with a random mutation probability. That is, $x_{ij}$ and $y_{ij}$ in \eqref{individual} change randomly in the range of $[0, L]$ and $[0, W]$.

\item[\bf S4)]\textbf{Evaluation:} Compute and update the fitness value of each individual again using (\ref{fitness}).

\item[\bf S5)]\textbf{Optimization:} If the current optimal fitness value is worse than that of the previous generation, the worst individual in the current generation is replaced with the best individual in the previous generation.

\item[\bf S6)]\textbf{Iteration:} $Iterator$ is initially set to $1$, and repeat the steps S1-S5 till the preset maximum number of iterations $It_{max}$ is achieved. Finally, the individual with the largest fitness value stored in the last position of the population array can be obtained.
\end{enumerate}

\subsection{PSO-based deployment}\label{}
PSO is a swarm intelligence algorithm designed by simulating the predation behavior of bird swarm. Each candidate solution is represented by one particle in the swarm. At first, all particles in swarm are assigned initial random positions and velocities. Then the position of each particle is updated in turn according to the velocity of each particle, the best global position known in the problem space, and the best known position of the particle. As the computation process advances, by exploring and leveraging known vantage points in the search space, particles gather or aggregate around one best particle (a.k.a. optimized solution).

In PSO-based transmitter deployment, each particle $\bm{Q}_i$ represents a possible deployment solution containing series random transmitter's coordinates, which can be described as
\begin{equation}\label{particle}
    \bm{Q}_i =\{\bm{q}_{ij}(x_{ij}, y_{ij})~|~1 \le i \le M_p, 1 \le j \le M_t, (x_{ij}, y_{ij})\in \bm{G}\}
\end{equation}
The swarm consists of $M_p$ particles, and $M_t$ is the number of transmitters contained in one particle. Same as GA-based deployment scheme, the optimal solution is a set of transmitters' coordinates maximizing the objective function \eqref{objectivesum} (i.e., the best particle). Moreover, the fitness function of the PSO-based scheme can be depicted as
\begin{equation}\label{particlefitness}
\begin{aligned}
    g(\bm{Q}_i) = \omega_1\times\frac{N_{covered}}{N_{total}} + \omega_2\times\frac{\sum_{i=1}^{N_{covered}}(1-S_c(\mathbf{r}_i))}{\sum_{i=1}^{N_{total}}(1-S_c(\mathbf{r}_i))}.
\end{aligned}
\end{equation}

Different from GA-based deployment scheme, each particle contains not only the position of each transmitter $\bm{q}_{ij}$, but the moving velocity $\bm{v}_{ij}$ when transmitter's position is updated. Furthermore, in PSO-based deployment, each transmitter's position is updated with reference to global best result $\bm{gbest}_{ij}$ (i.e., the global best position experienced by every particle in the particle swarm after comparing with each other) and local best result $\bm{pbest}_{ij}$ (i.e., the best position each particle has ever experienced itself) during the iteration. For $(k+1)th$ particle movement, each particle's velocity will be updated as follow:
\begin{equation}\label{velocityupdate}
\begin{aligned} \bm{v}_{i j}^{k+1}=& \omega \bm{v}_{i j}^{k}+c_{1}\operatorname{rand}_{1}\left(\bm{pbest}_{ij}-\bm{q}_{ij}^{k}\right)+\\ & c_{2} \operatorname{rand}_{2}\left(\bm{gbest}_{ij}-\bm{q}_{ij}^{k}\right) \end{aligned}
\end{equation}
where $i = 1,2,..., M_p,$ and $j = 1,2,..., M_t.$ $\omega$ is called ``inertial weight'' which determines the possibility that the particle will stay on its current trajectory. $c_1$ and $c_2$ represent the scale that the particle will move toward $\bm{pbest}_{ij}$ and $\bm{gbest}_{ij}$, respectively. $\operatorname{rand}_{1}$ and $\operatorname{rand}_{2}$ are random numbers uniformly distributed in $[0,1]$. Then, each particle's position will be updated as
\begin{equation}\label{positionupdate}
\bm{q}_{i j}^{k+1}=\bm{q}_{ij}^{k}+\bm{v}_{ij}^{k+1}
\end{equation}

The optimal RBC transmitter deployment through PSO is depicted in Algorithm 2, and the steps are as follows:

\begin{algorithm}[!h]
    \caption{PSO-based deployment}
    \begin{algorithmic}[1]
    \Require $L$, W, $r$;
    \State \text{Initialization:}
    \State $\text{\  Initializing\ each\ particle's\ position\ and\ velocity}$;
    \State $\text{\  Initializing\ parameters\ of\ PSO}$;
    \While {$Iterator \le It_{max}$}
    \State $\text{\ Compute\ each\ particle's\ fitness\ with\ \eqref{particlefitness}}$ \ (cf. T1);
    \State $\text{\ Find\ the}$ $\bm{pbest}$ $\text{\ and}$ $\bm{gbest}$ (cf. T2);
    \State $\text{\ Update\ particle's\ velocity\ and\ position\ }$\ (cf. T3);
    \EndWhile
    \State \Return{$\text{The\ best\ particle}$}.
    \label{TDGA}
    \end{algorithmic}
    \end{algorithm}

\begin{figure}[htbp]
	\centering
    \subfigure[GA-based deployment]{
    \includegraphics[width=0.35\textwidth]{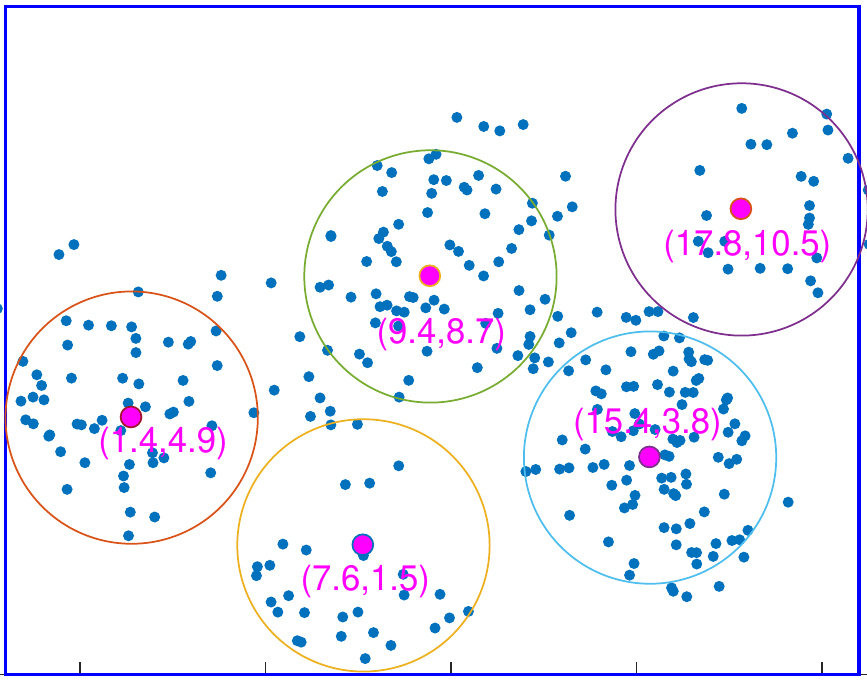}
    \label{deploymentGA}
    }
    \subfigure[PSO-based deployment]{
    \includegraphics[width=0.35\textwidth]{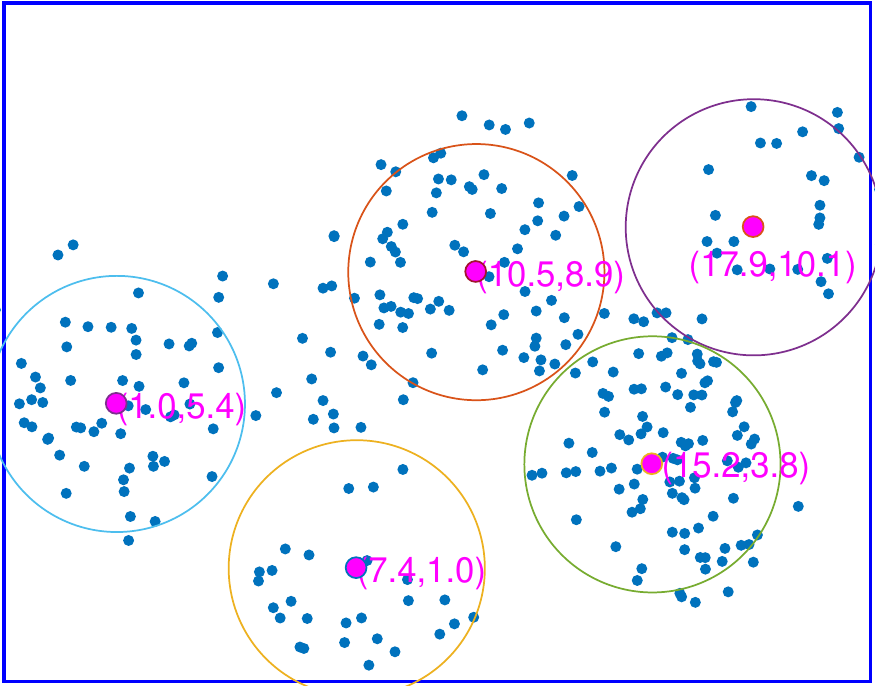}
    \label{deploymentPSO}
    }

    \subfigure[Uniform deployment]{
    \includegraphics[width=0.35\textwidth]{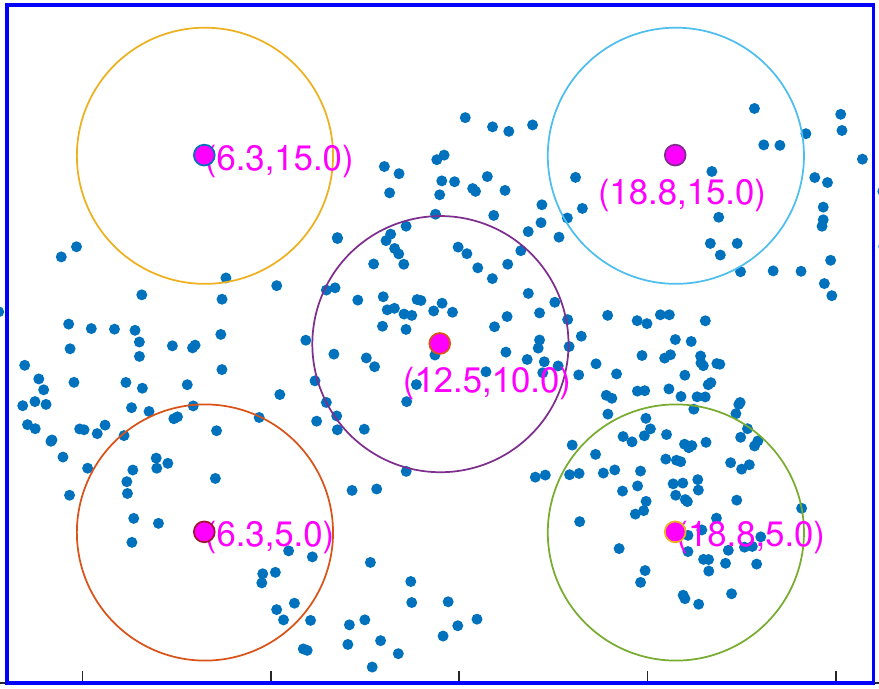}
    \label{deploymentGrid}
    }
    \subfigure[Random deployment]{
    \includegraphics[width=0.35\textwidth]{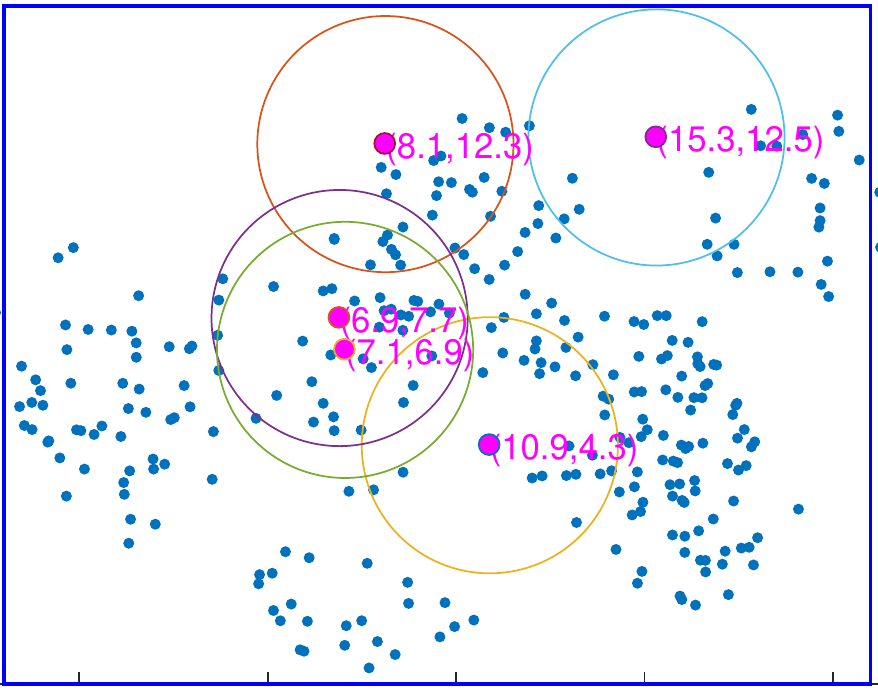}
    \label{deploymentRand}
    }

    \caption{ Deployment implementation of $5$ transmitters at a height of $5$m in a $25$m$\times 20$m rectangular region.}
    \label{deployment}
\end{figure}

\begin{enumerate}
\item[\bf T1)]\textbf{Evaluation:} Compute and update the fitness value of each particle using (\ref{particlefitness}).

\item[\bf T2)]\textbf{Finding the best:} For each particle, compare the fitness value computed through \eqref{particlefitness} of its current position with that of the best position it passed, and update the local best position $\bm{pbest}$. Meanwhile, compare fitness value of each particle's best position it passed with that of the best position in the whole swarm, and update the global optimal position $\bm{gbest}$.

\item[\bf T3)]\textbf{Update:} Update each particle's velocity and position with \eqref{velocityupdate} and \eqref{positionupdate}, respectively.

\item[\bf T4)]\textbf{Iteration:} The iteration process is consistent with GA-based algorithm. Consequently, the particle with the largest fitness value stored in the last position of the swarm array can be obtained.
\end{enumerate}

GA-based and PSO-based deployment schemes aim at finding a set of transmitters' coordinates to realize the optimal covering of receivers with least transmitters for high charging efficiency. Given the region's size and the RBC's covering range, the optimized deployment scheme with each transmitters' coordinate can be achieved.

\section{Performance Analysis}\label{Section5}
In this section, we evaluate performances of the proposed GA-based and PSO-based RBC deployment schemes in terms of the following two aspects:
1) quality of coverage under a varying number of transmitters with the GA-based, PSO-based, uniform, and random deployment method;
2) charging efficiency improvement analysis of the GA-based and PSO-based deployment method under a certain transmitter number, with the comparison of the uniform and random transmitter placement method in a given region.

We considered two settings, in which the RBC transmitters are to be placed at a height of $3$m and $5$m, respectively. With regards to the charging service coverage of an RBC transmitter computed by \eqref{coveringmodel}, as the power required for charging mobile / IoT devices e.g., mobile phone is $5$W \cite{ChargePower2014}, we define the RBC service coverage to be the area within which the received power of every device is at least $5$W. By taking $P_{min}=5$W in \eqref{coveringmodel}, we find that the charging service covering range of an RBC transmitter placed at a height of $3$m ($5$m) with input power of $200$W is $5.2425$m ($3.3888$m).

We considered a square area with a length of $25$m, and a rectangular area with a length of $25$m and a width of $20$m. In receiver distribution simulation, $\lambda$ was set to $9$, and $\sigma$ to 3. In GA-based deployment, the population size was set to $100$, the maximum number of generations to $5000$, the mutation probability $p_m$ to $0.15$, the crossover probability $p_c$ to $0.8$, and $\omega_1 = \omega_2 = 0.5$. In PSO-based deployment, the swarm size was set to $100$, the maximum number of generations to $5000$, positive constants $c_1$ and $c_2$ that the old velocity moves towards $\bm{pbest}$ and $\bm{gbest}$  to $2$, and inertial weight $\omega$ will change linearly from $0.9$ to $0.2$ during the iteration of algorithm. The maximum number of iterations was $5,000$. All Simulations are performed using MATLAB.

\subsection{Receiver distribution and transmitter deployment}
To demonstrate the superiority of the proposed deployment algorithm to deploy RBC transmitters for high-efficiency charging service, we compare the GA-based and PSO-based algorithm with both the uniform and random deployment method. The random deployment denotes that each transmitter's position is generated randomly within a given region. The uniform deployment means that the transmitters are placed evenly both horizontal and vertical, dividing the length and width of a given area. The simplified demonstration of the uniform deployment process which we used in the simulation can be described as follows:

1) Given the length $L$ and the width $W$ of the region, and the transmitter number $N_t$, if $n^2 < N_t \le (n+1)^2,~n\in \bm{N}_+$, the region will be divided into $(n+1)^2$ squares, while the length and width contains $(n+1)$ squares, respectively.

2) Then, if $(n+1)$ is odd, place the first transmitter in the central square of the region, and other transmitters are put in the squares spreading out from the central square in turn; otherwise, the transmitters are placed first in the innermost squares on the diagonals, and other transmitters are put in the squares spreading out from them.

We present GA-based deployment and uniform deployment of $5$ transmitters at a height of $5$m in a $25$m$\times20$m region as examples, respectively. In this setting, the receiver number is $300$, of which the distribution obeys Thomas distribution. Figure \ref{deploymentGA} depicts the GA-based deployment implementation of $5$ transmitters. Scatter points represent the RBC receivers. The rectangular area in blue is the target region to be covered, whose length is $25$m and width $20$m. Each circle stands for the coverage of a single transmitter positioned at a height of $5$m. The position of each circle's center marks the position coordinates of each RBC transmitter. PSO-based deployment can be seen from Fig. \ref{deploymentPSO}.
It's evident that the receiver distribution is non-uniform, and the deployment results obtained by the optimized GA-based and PSO-based schemes are similar, which can both cover the area with more receivers.

Figure \ref{deploymentGrid} depicts the uniform deployment implementation, in which the RBC transmitters are distributed evenly both on horizontal and vertical lines. Although uniform deployment is simple and practical, it is obvious that uniform deployment does not cover as many devices to be charged as possible, resulting in resource wastes. The random deployment case is shown in Fig. \ref{deploymentRand}, where the large overlap of transmitters' coverage occurs. Hence, the deployment may be farther away from the desired optimization deployment effect.

Overall, the uniform deployment can deploy the transmitters uniformly and regularly, thus avoiding the difficulty of wiring and facilitating deployment. However, the GA-based and PSO-based algorithms can cover more receivers with the same number of transmitters through a computationally more involved procedure, thus to improve the charging efficiency and quality in a given area.

\subsection{Quality of coverage}

\begin{figure}[!t]
	\centering
    \includegraphics[scale=0.6]{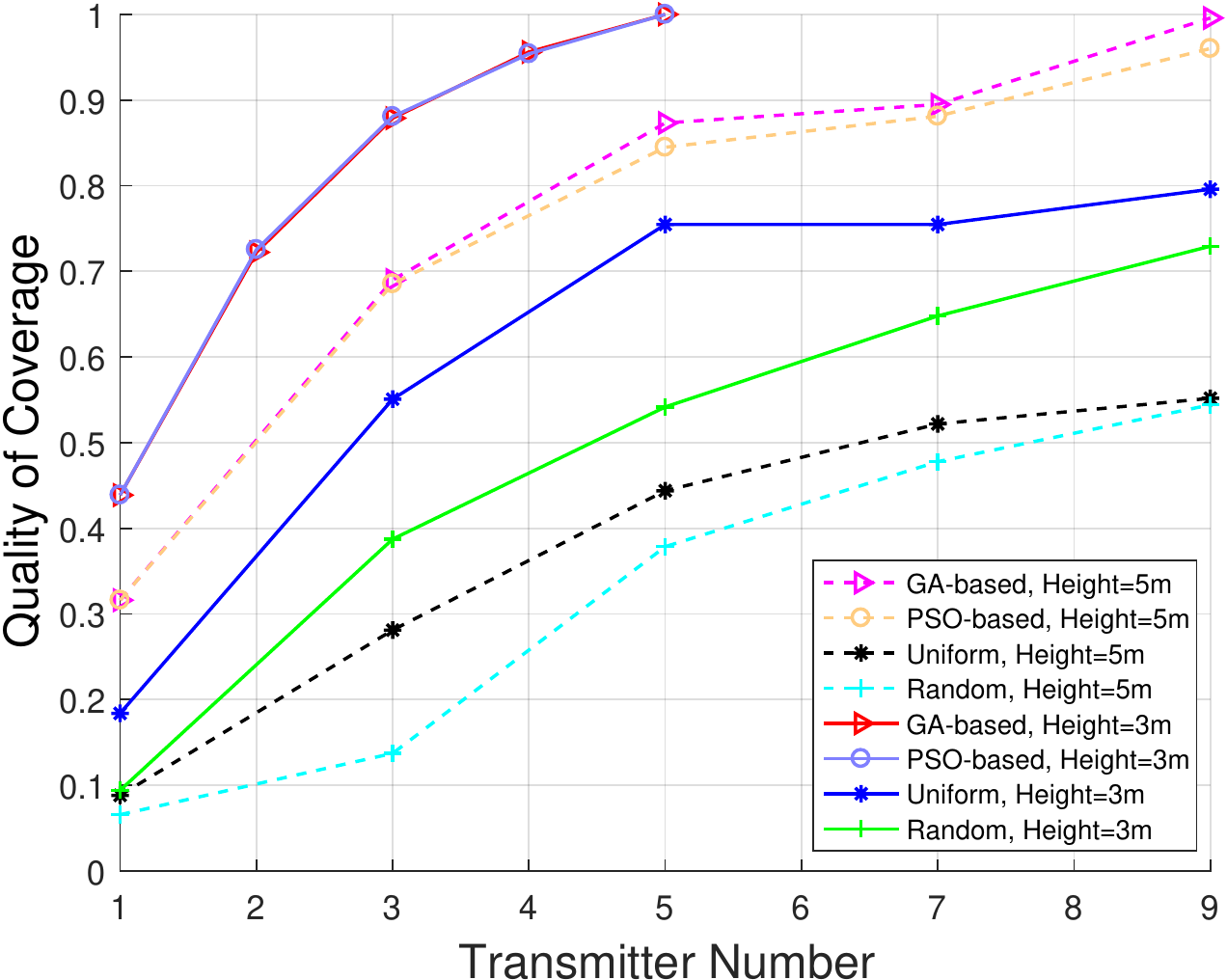}
	\caption{Quality of coverage as a function of the number of deployed transmitters placed at a height of $3$m and $5$m in a $25$m$\times 20$m rectangular region.}
    \label{qoc}
    \end{figure}
Consider deploying the RBC transmitters at a height of $3$m and $5$m in a $25$m$\times 20$m rectangular region using both proposed and uniform algorithms, while the receiver number is 300 obeying Thomas distribution. We define the coverage quality of the proposed algorithm according to the fitness function \eqref{fitness} and \eqref{particlefitness}. Figure \ref{qoc} shows how the quality of coverage improves as the transmitter number increases for the GA-based, PSO-based, uniform, and random deployment schemes, with the transmitters deployed at a height of $3$m and $5$m, respectively. With the transmitter deployment at a height of $3$m, $5$ transmitters are required to reach the maximum coverage quality with both GA-based and PSO-based deployment method, while the coverage quality with uniform deployment keeps lower than that with the proposed method, and remains higher than that with random deployment. Evidently, GA-based is slightly better than PSO-based deployment algorithm.

The experiments are repeated for transmitters placed at a height of $5$m. For obtaining a best coverage of quality, one may need at least $9$ transmitters. While the coverage quality is less than 0.6 with the same number of transmitters under the uniform and random deployment. Evidently, we can derive that to reach a high coverage quality, deploying transmitters at a height of $3$m use less transmitters than placing them at a height of $5$m. Thus, if the transmitter number is limited, we can reduce the RBC-embedded UAV's flight height to achieve high quality of coverage.

\begin{figure}[!t]
	\centering
    \includegraphics[scale=0.6]{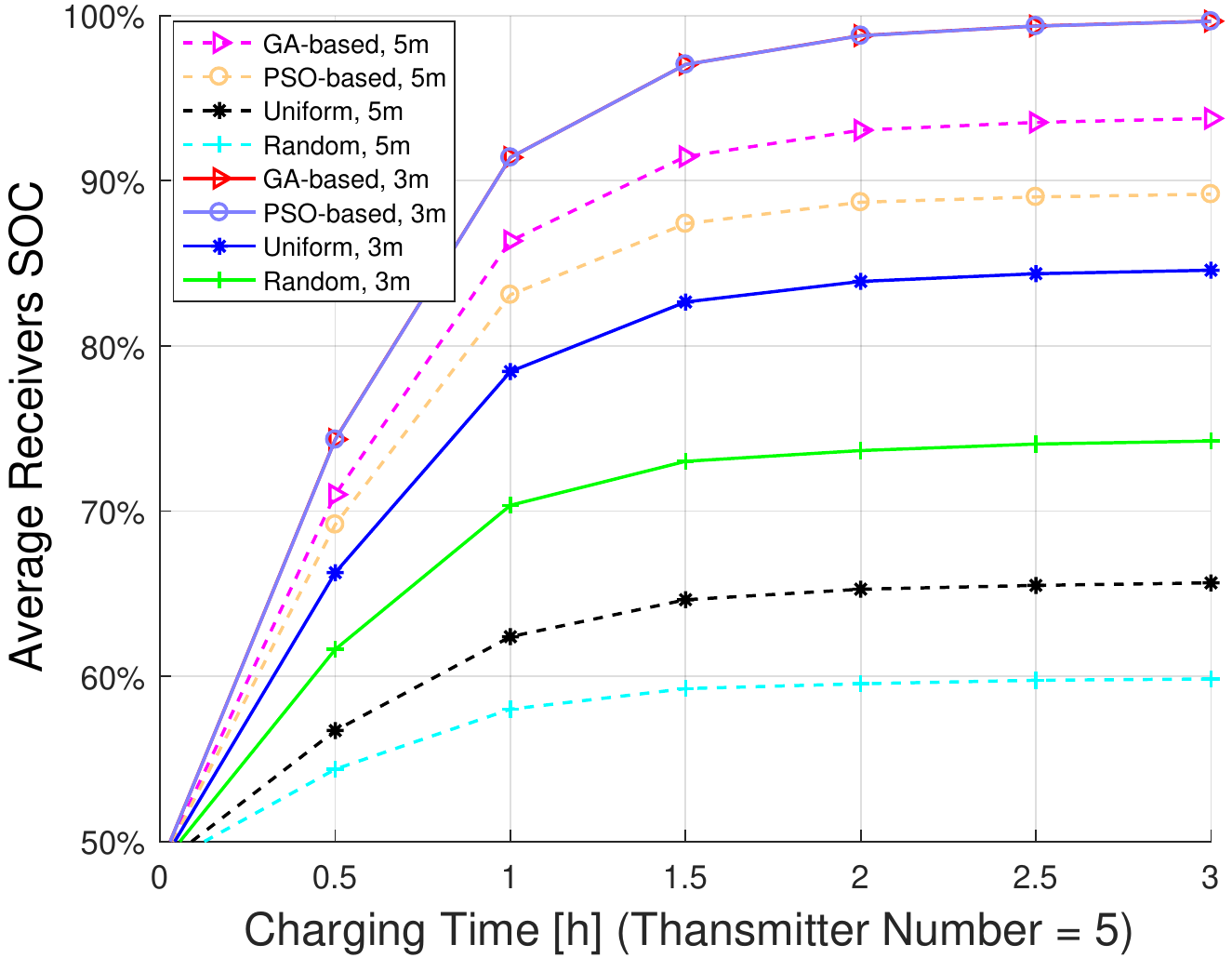}
	\caption{Average receivers SOC in terms of charging time of $5$ transmitters with GA-based, PSO-based, uniform, and random algorithm at a height of $3$m and $5$m in a $25$m$\times 25$m region.}
    \label{chargingtime625}
    \end{figure}

\begin{figure}[!t]
	\centering
    \includegraphics[scale=0.6]{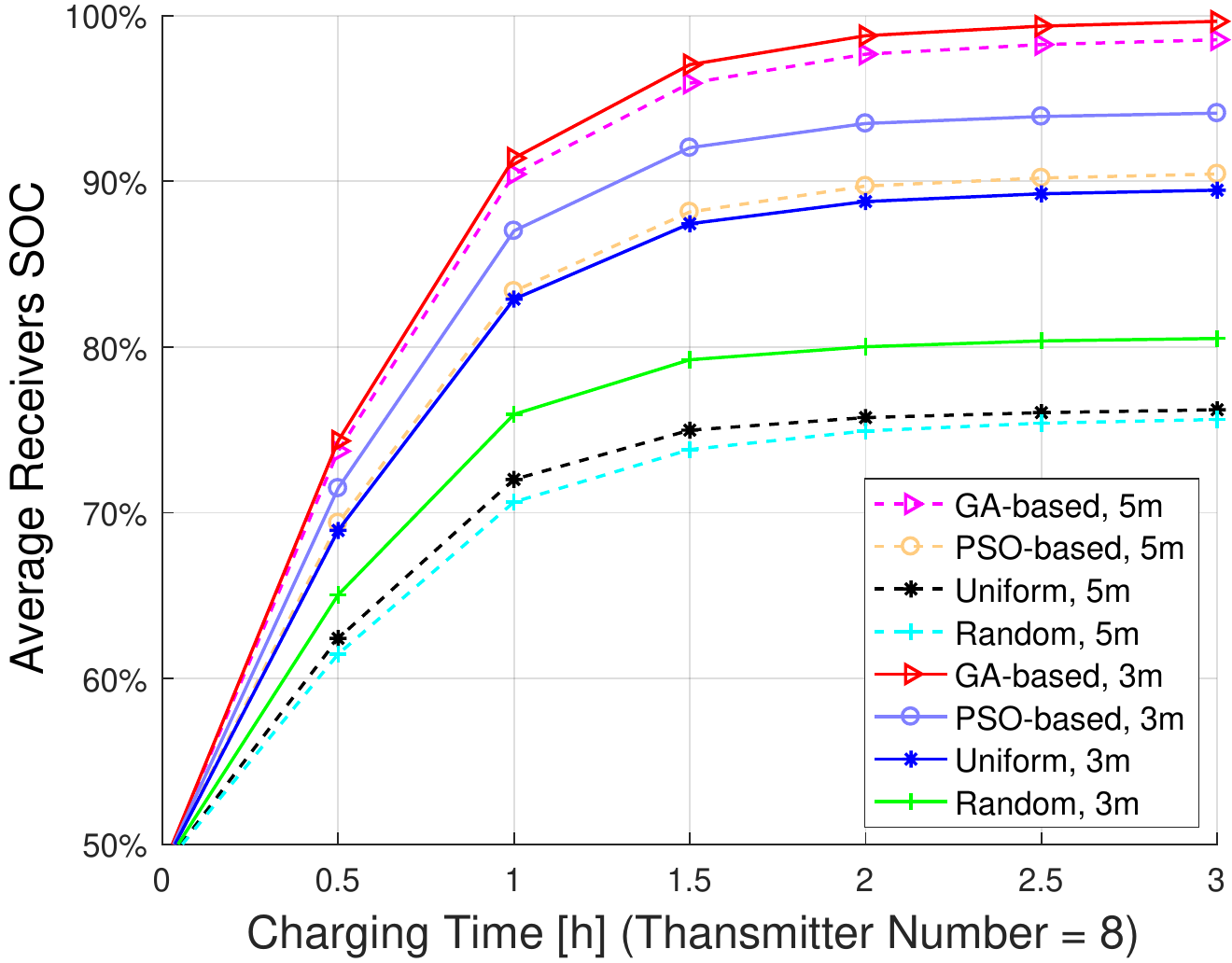}
	\caption{Average receivers SOC in terms of charging time of $8$ transmitters with GA-based, PSO-based, uniform, and random algorithm at a height of $3$m and $5$m in a $25$m$\times 25$m region.}
    \label{chargingtime625t8}
    \end{figure}

\subsection{Charging efficiency}

In order to evaluate the charging performance of the proposed deployment algorithm for the receivers in an area, we did the following comparative experiments: 1) based on the charging and discharging model, the receivers are charged according to the scheduling rule formulated in Sec.~\ref{Evaluation}; 2) the average receivers State of Charge (SOC), i.e., remaining
capacity percentage of \eqref{SOC} after charging for some times under the GA-based and PSO-based deployment are compared with that under both the uniform and random deployment; 3) the average receivers SOC after charging for certain times with various transmitter and receiver number under the proposed transmitter deployments are compared with that under uniform and random deployment.

Figure~\ref{chargingtime625} shows average receivers SOC in terms of charging time with $5$ transmitters placed at a height of $3$m
and $5$m under the GA-based, PSO-based, uniform, and random deployment, respectively. If the transmitters are placed at a height of $3$m, the average receivers SOC under GA-based deployment is extremely slight higher than that under PSO-based deployment, 15\% higher than uniform deployment, and 25\% higher than random deployment; while if the transmitters are placed at a height of $5$m, the average receivers SOC under GA-based deployment is 5\%, 30\%, and 35\% higher. Obviously, the results show that GA-based deployment scheme outperforms the PSO-based, uniform, and random deployment in terms of minimizing the charging time for increasing the average receivers SOC to a certain level.

\begin{figure}[!t]
	\centering
    \includegraphics[scale=0.6]{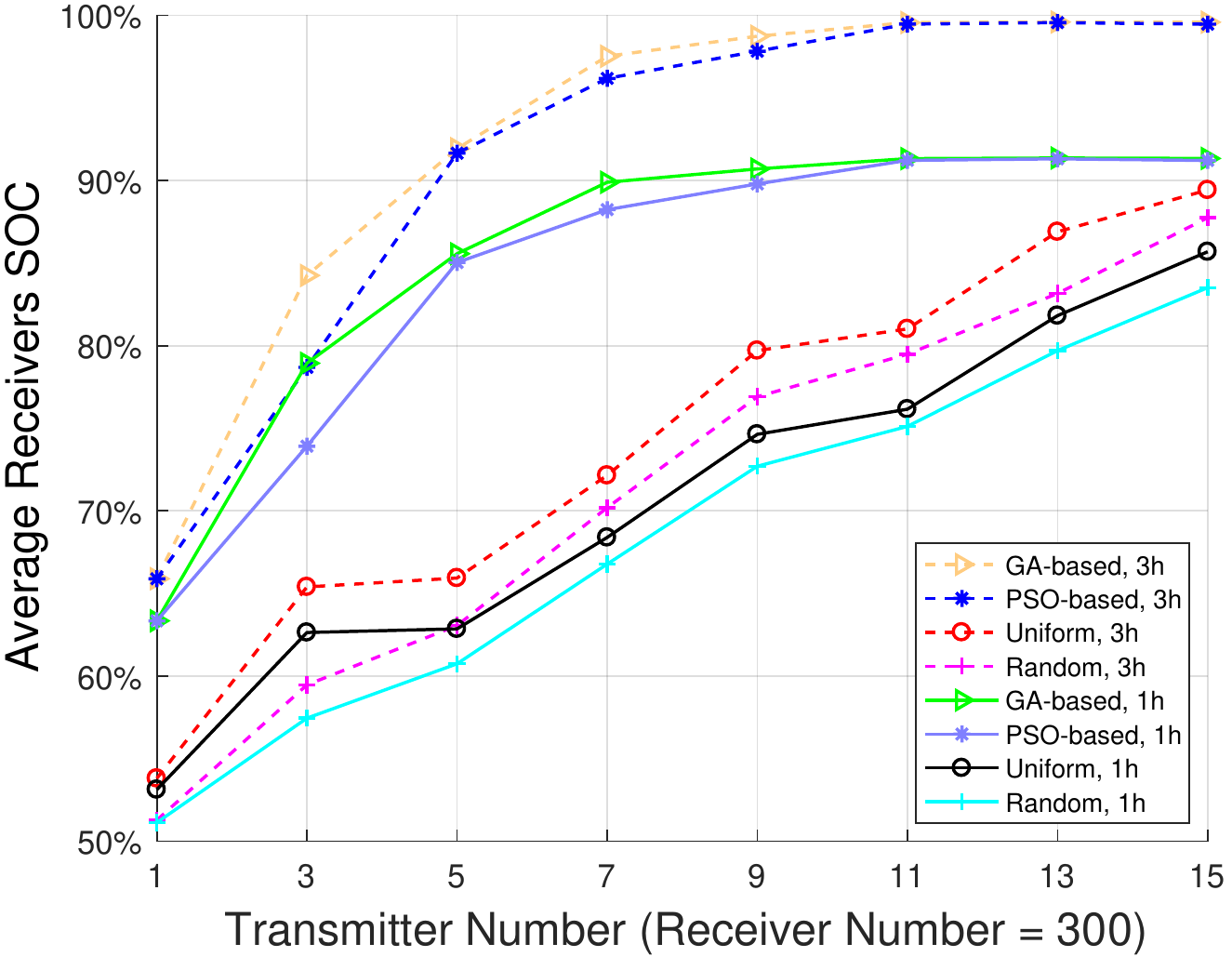}
	\caption{Average receivers SOC in terms of transmitter number with $300$ receivers under GA-based, PSO-based, uniform, and random deployment after charging for $1,3$h at a height of $5$m in a $25$m$\times 25$m region.}
    \label{transmitternumber}
    \end{figure}

According to Fig. \ref{chargingtime625t8}, same experiments have been done with $8$ RBC transmitters. When the transmitters are placed at the hight of $3$m, the GA-based deployment outperforms PSO-based, uniform, and random deployment in charging efficiency improvement for around 6\%, 10\%, and 20\%, while the GA-based deployment can improve the average receivers SOC 9\%, 22\%, and 23\% more than PSO-based, uniform, and random deployment with the transmitters deployed at a height of $5$m. We can also derive the conclusion that as the number of transmitters increases, the superiority of the optimized GA-based and PSO-based algorithms decrease.

    \begin{figure}[!t]
	\centering
    \includegraphics[scale=0.6]{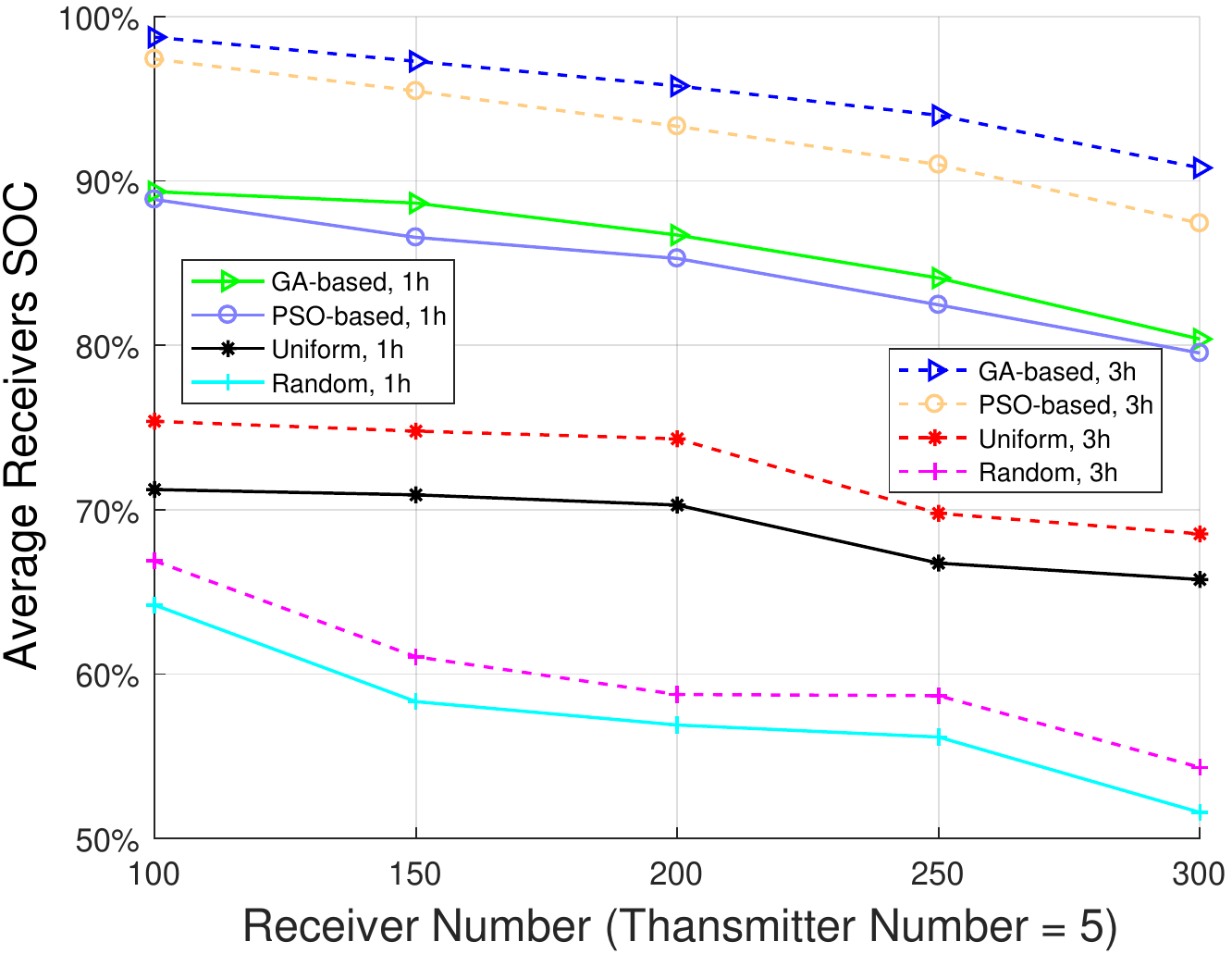}
	\caption{Average receivers SOC in terms of receiver number with $5$ transmitters under GA-based, PSO-based, uniform, and random deployment after charging for $1,3$h at a height of $5$m in a $25$m$\times 25$m region.}
    \label{receivernumber}
    \end{figure}

We present average receivers SOC in terms of transmitter number with $300$ receivers under GA-based, PSO-based, uniform, and random deployment after charging for $1,3$h at a height of $5$m in a $25$m$\times 25$m rectangular region. As shown in Fig.~\ref{transmitternumber}, the GA-based and PSO-based algorithm performs similarly, and outperforms uniform and random deployment under various transmitter number. Moreover, as the number of transmitter increases, the superiorities on charging efficiency of the GA-based and PSO-based method
increase first and then decreases, with the comparison of both the uniform and random deployment method. That is, if the transmitter number is too small or too large, the performance of the four deployment methods will not differ too much.

Besides, we did the experiments to depict how the receiver number in a given region effects the charging efficiency. Plots in Fig.~\ref{receivernumber} demonstrate the average receivers SOC change in terms of receiver number with $5$ transmitters after charging for $1, 3$h. The simulations have been done under GA-based, PSO-based, uniform, and random deployment with the transmitters at a height of $5$m in a $25$m$\times 25$m square region. Obviously, the GA-based and PSO-based deployment outperform the uniform and random deployment in increasing average receivers SOC under $1, 3$h charging. Furthermore, the performance of uniform deployment is better than random deployment. However, with the increase of the receiver number in a given region, the average receivers SOC presents the downtrend. It is intuitive as more receivers needs more transmitters to supply enough power for improving the average SOC of receivers.

\section{Conclusions}\label{Section6}

This paper advocated the optimized wireless power transmitter deployment scheme to offer wireless charging services for multi-user in a given region, balancing the charging fairness and charging service quality. RBC, as a long-range, high-power, and safe WET technology, can provide WiFi-like wireless charging services. At first, we introduced the features and energy attenuation principle of the RBC system. Then, the coverage model of an RBC transmitter was proposed. After introducing the RBC receiver charging / discharging model and the scheduling charging rule, we presented a GA-based and a PSO-based transmitter deployment scheme for providing high-efficiency and high-quality wireless charging services in a certain area. The proposed algorithms outperform the uniform and random deployment in coverage quality and maximizing the average receivers capacity after charging for a certain time.

 There are several exciting directions for future research, including: 1) optimizing the GA and PSO to improve the accuracy and effectiveness; and, 2) studying RBC transmitter deployment in dynamic wireless power networks; and, 3) accounting for the costs, user capacity, and multi-tier deployment.

\bibliographystyle{IEEEtran}
\bibliographystyle{unsrt}
\bibliography{references}

\end{document}